\documentclass[prd,nofootinbib,twocolumn]{revtex4}
\usepackage{amsmath}
\usepackage{graphics}
\usepackage{epsfig}
\begin{document}
\title{$\rho(\omega) \to \pi^0 \pi^0 \gamma$, $\rho(\omega)
\to \eta \pi^0 \gamma$ decays in the local quark Nambu--Jona-Lasinio model}

\author{A. E. Radzhabov}%
\email{aradzh@theor.jinr.ru}
\author{M. K. Volkov}
\email{volkov@theor.jinr.ru}
\author{N. G. Kornakov}%
\email{kornakov@theor.jinr.ru}

\affiliation{%
Bogoliubov Laboratory of Theoretical Physics, Joint Institute for Nuclear Research,
141980 Dubna, Russia }

\date{\today}

\begin{abstract}
The branching ratios and photon spectra of the rare processes $\rho(\omega) \to \pi^0 \pi^0
\gamma$, $\rho(\omega) \to \eta \pi^0 \gamma$ are calculated in the framework of the standard
local quark Nambu--Jona-Lasinio (NJL) model. Three types of diagrams are considered: the quark
box and the pole diagrams with scalar ($\sigma,a_0(980)$) and vector ($\rho,\omega$) mesons.
The obtained estimations for the widths of the processes $\rho(\omega) \to \pi^0 \pi^0 \gamma$
are in satisfactory agreement with existing experimental data. Predictions are made for the
widths of the processes $\rho(\omega) \to \eta \pi^0 \gamma$.

\end{abstract}
\maketitle
\section{Introduction}

At the present time, there are many theoretical and experimental works devoted to
investigation of the rare decay processes $\rho(\omega)\to \pi^0(\eta) \pi^0 \gamma$. These
processes are very interesting for studying the mechanism of chiral symmetry breaking of
strong interaction of hadrons.

Recently, the branching ratios for the rare decays of $\rho$ and $\omega$ mesons to the pair
of pions have been measured with good accuracy by SND \cite{Achasov:2002jv} and CMD-2
\cite{Akhmetshin:2003rg} collaborations at VEPP-2M $e^+e^-$ collider (see table
\ref{tab:ropipiga} where theoretical estimations and experimental results are given). The
situation for decays with $\eta$ meson in the final state is worse than for radiative two pion
production, there is an estimation for the upper limit for the decay of $\omega$-meson
$\mathcal{B}(\omega\rightarrow\eta\pi^0\gamma)<3.3\times 10^{-5}$ at 90\% C.L. only
\cite{Akhmetshin:2003rg}. Therefore, the theoretical predictions for the decays of vector
mesons into $\eta\pi^0\gamma$ are of large interest.

Indeed, there are many theoretical works devoted to the description of these processes in
different models. For instance, in one of the first works devoted to the calculations of these
decays, the vector meson dominance (VMD) model was used \cite{Singer:1962qr,Singer:1963ae}.
Only diagrams with intermediate vector mesons were taken into account and only rough
estimations of these processes were obtained. In 1992, in the framework of a similar model
these processes were estimated more accurately and it was found that the VMD leads to too low
values of the branching ratios \cite{Bramon:1992kr} (see table 1). Lately in
\cite{Palomar:2001vg,Gokalp:2000xy,Gokalp:2000ir,Bramon:2001un} it was found that inclusion of
diagrams with scalar meson exchange increases branching ratios, which leads to better
agreement with experiment. There are different methods to take into account the effect of
scalar meson exchange. One of them is connected with phenomenological inclusion of the scalar
pole diagram with the mass and the width fixed from experiment
\cite{Gokalp:2000xy,Gokalp:2000ir,Kucukarslan:2004zc}. This method of inclusion of the scalar
meson leads to the breaking of chiral symmetry. Other methods taking into account scalar
effects are connected with dynamical generation of the scalar meson after unitarization of
pseudoscalar meson loop diagrams \cite{Palomar:2001vg} or based on linear sigma model
\cite{Bramon:2001un,Escribano:2006mb}. Let us note that in most these models it was necessary
to use additional parameters for description of the above-mentioned rare decays of vector
mesons.

\begin{table}
\caption{\label{tab:ropipiga} Branching ratios for the processes $
\rho\rightarrow\pi^0\pi^0\gamma$ and $\omega\rightarrow\pi^0\pi^0\gamma$ obtained in
experiment (upper part) and theoretically(lower part). }
\begin{tabular}{|c|c|c|c|c|c|c|c|c|c|}
 \hline
          &$ \rho\rightarrow\pi^0\pi^0\gamma$, $10^{-5}$& $ \omega\rightarrow\pi^0\pi^0\gamma$, $10^{-5}$  \\
 \hline
 SND \cite{Achasov:2002jv}
  &$ 4.1^{+1.0}_{-0.9}\pm 0.3 $& $ 6.6^{+1.4}_{-1.3}\pm 0.6  $  \\
 CMD-2 \cite{Akhmetshin:2003rg}
 &$ 5.2^{+1.5}_{-1.3}\pm 0.6 $& $ 6.4^{+2.4}_{-2.0}\pm 0.8  $  \\
 PDG \cite{PDBook}
            &$ 4.4\pm 0.8  $& $ 6.7\pm 1.1  $  \\
 \hline
\cite{Bramon:1992kr}    &$ 1.1  $&$ 2.8  $        \\
\cite{Palomar:2001vg}   &$ 4.2  $&$ 4.7  $        \\
\cite{Bramon:2001un}    &$ 3.8  $&$ 4.5 \pm 1.1 $ \\
\cite{Escribano:2006mb} &$ 4.2  $&$ 3.5-4.6$\\
This work               &$ 4    $&$ 8.3$\\
 \hline
\end{tabular}
\end{table}

\begin{table}
\caption{Branching ratios for the processes $ \rho\rightarrow\eta \pi^0\gamma$ and
$\omega\rightarrow\eta \pi^0\gamma$ obtained theoretically. }
\begin{tabular}{|c|c|c|c|c|c|c|c|c|c|}
 \hline
          &$ \rho\rightarrow\eta \pi^0\gamma  $& $ \omega\rightarrow\eta \pi^0\gamma  $  \\
 \hline
\cite{Bramon:1992kr}     &$ 4  \times 10^{-10}  $&$ 1.6\times 10^{-7}   $        \\
\cite{Palomar:2001vg}    &$ 7.5\times 10^{-10}  $&$ 3.3\times 10^{-7}  $        \\
\cite{Kucukarslan:2004zc}&$ 2.3\times 10^{-8}  $&$ 5.72\times 10^{-7}  $        \\
\cite{Escribano:2006mb}  &$ 5.2\times 10^{-10}  $&$ 9.7\times 10^{-8}$\\
This work                &$ 1.64\times 10^{-9}    $&$ 3.65\times 10^{-7}$\\
 \hline
\end{tabular}
\end{table}
In the present paper, we suggest using the well-known $U(3)\times U(3)$ local quark NJL model
\cite{Nambu:1961tp,Eguchi:1976iz,Kikkawa:1976fe,Volkov:1982zx,Ebert:1982pk,Volkov:1984kq,Volkov:1986ec,Ebert:1985kz,Klimt:1989pm,Klevansky:1992qe,Volkov:1993jw,Hatsuda:1994pi,Ebert:1994mf,Volkov:1998ax,Volkov:2006vq}
for the description of decays of $\rho$-,$\omega$-mesons into a pair of neutral pseudoscalar
mesons. The advantage of the model is that for descriptions of these processes it is not
necessary to introduce any additional parameter. Three types of diagrams are taken into
account: quark box and pole diagrams with intermediate scalar ($\sigma,a_0(980)$) and vector
($\rho,\omega$) mesons. In addition, the estimations are given for the processes whose
vertices appear in the amplitudes of pole diagrams.

The obtained results for the decay processes $\rho(\omega) \to \pi^0 \pi^0 \gamma$ are in
satisfactory agreement with experimental data. Predictions for decay processes $\rho(\omega)
\to \eta \pi^0 \gamma$ do not contradict existing experiments.

\section{Model parameters and description of intermediate processes}

The $U(3)\times U(3)$ NJL model is based on the effective four-quark interaction in the
scalar--pseudoscalar and vector--axial-vector channels and six-quark t`Hooft interaction.
The weak interactions are introduced in the model with the help of redefining a kinetic
quark term and electromagnetic interactions are introduced with help of VMD. Since the
details of the model are described in many papers (see e.g.
\cite{Volkov:1986ec,Volkov:1993jw,Ebert:1994mf,Volkov:1998ax,Volkov:2006vq}), we give
here only general characteristics of this model and parameters of the model which are
needed for our calculations. The model contains the mechanism of spontaneous breaking of
chiral symmetry and has six arbitrary parameters: constituent masses of nonstrange
$m_u=m_d=263$ MeV and strange $m_s=407$ MeV quarks, $O(4)$ cut-off $\Lambda=1.27$ GeV,
the four-quark coupling constants in scalar--pseudoscalar $G_1=4.16$ GeV$^{-2}$ and
vector--axial-vector $G_2=14.66$ GeV$^{-2}$ channels, and the six-quark t`Hooft coupling
constant $K=12.5$ GeV$^{-5}$. The model parameters are defined as in
\cite{Volkov:1998ax,Volkov:2006vq}\footnote{Note that in the present paper for
definition of model parameters we have used recent experimental data \cite{PDBook} for
$\rho$-meson mass and decays $\pi\to \mu\nu$, $\rho\to \pi\pi$.} using experimental
values of masses of the pion, kaon and $\rho$-meson, $\eta-\eta^\prime$ mass difference,
the weak pion decay constant $f_\pi=92.4$ MeV, and the strong decay width $\rho\to
\pi\pi$ ($g_\rho=5.94$). As a result, in the NJL model it is possible to describe the
mass spectrum and main strong and electroweak decays of pseudoscalar, scalar, vector,
and axial-vector meson nonets
\cite{Volkov:1986ec,Ebert:1985kz,Klimt:1989pm,Klevansky:1992qe,Volkov:1993jw,Hatsuda:1994pi,Ebert:1994mf,Volkov:1998ax,Volkov:2006vq,Volkov:1999yi}.

\begin{table}
\caption{\label{tab:mass} Meson masses obtained in the model in comparison with
experimental data \cite{PDBook}.}
\begin{tabular}{|c|c|c|c|c|c|c|c|c|c|}
 \hline
          &model& experiment  \\
 \hline
$ M_\pi          $ &$ 135  $&$ 134.9766 \pm 0.0006 $    \\
$ M_K            $ &$ 495  $&$ 497.648 \pm 0.022   $ \\
$ M_\eta         $ &$ 511  $&$ 547.51\pm 0.18 $\\
$ M_{\eta^\prime}$ &$ 972  $&$ 957.78\pm0.14  $ \\
$ M_\rho         $ &$ 775.5  $&$ 775.5\pm 0.4 $\\
$ M_{\sigma}     $ &$ 516  $&$ 400-1200       $ \\
$ M_{a_0}(980)   $ &$ 785  $&$ 984.7\pm 1.2   $ \\
$ M_{f_0}(980)   $ &$ 1100 $&$ 980  \pm 10    $\\
 \hline
\end{tabular}
\end{table}

In the following we need a meson-quark coupling constant. The scalar meson coupling constant
takes the form \cite{Kikkawa:1976fe,Volkov:1986ec,Volkov:2006vq}
\begin{eqnarray}
g_{a_0}=g_{\sigma_u} =[4I^{\Lambda}_2(m_u)]^{-1} = \frac{g_{\rho}}{\sqrt{6}},
\end{eqnarray}
where $I^{\Lambda}_2(m)$ is logarithmically
divergent integral
\begin{eqnarray}
I^{\Lambda}_2(m)={N_c\over (2\pi)^4}\int d^4_e k {\theta (\Lambda^2 -k^2)
\over (k^2 + m^2)^2} .
\label{DefI}
\end{eqnarray}

After taking into account the $\pi - a_1$ transition there is additional renormalization of
pion fields
\begin{eqnarray}
g_{\pi}=g_{\eta_u}=g_{a_0}Z^{1/2},\quad Z=\left(1-\frac{6m_u^2}{M_{a_1}^2}\right)^{-1}.
\end{eqnarray}
From the weak pion decay $\pi \to \mu \nu$ the Goldberger-Treiman relation follows
$f_\pi={m}/{g_{\pi}}$. Physical isoscalar mesons are the mixed states of the pure nonstrange
and strange states and t`Hooft interaction allows us to correctly describe this mixing. We
define these states as
\begin{eqnarray}
\eta       &=& -\eta_u   \sin{\bar \theta}+ \eta_s \cos{\bar \theta} , \nonumber \\
\eta^\prime&=&  \eta_u   \cos{\bar \theta}+ \eta_s \sin {\bar \theta},~~~{\bar \theta}=\theta - \theta_0\nonumber \\
\sigma     &=&  \sigma_u \cos{\bar \phi} - \sigma_s\sin {\bar \phi} , \nonumber \\
f_0        &=&  \sigma_u \sin{\bar \phi} + \sigma_s\cos {\bar \phi} ,~~~{\bar \phi} = \theta_0
- \phi \label{mix}
\end{eqnarray}
where $\theta_0 \approx 35.3^{\circ}$ is the ideal mixing angle $({\rm ctg}~ \theta_0={\sqrt
2})$, $\theta=-18.1$ and $\phi=23.4$ are the singlet-octet mixing angles for pseudoscalar and
scalar mesons \cite{Volkov:1998ax}. Meson masses obtained in the model are given in table
\ref{tab:mass}.

\begin{figure*}
\resizebox{17.8cm}{!}{\includegraphics{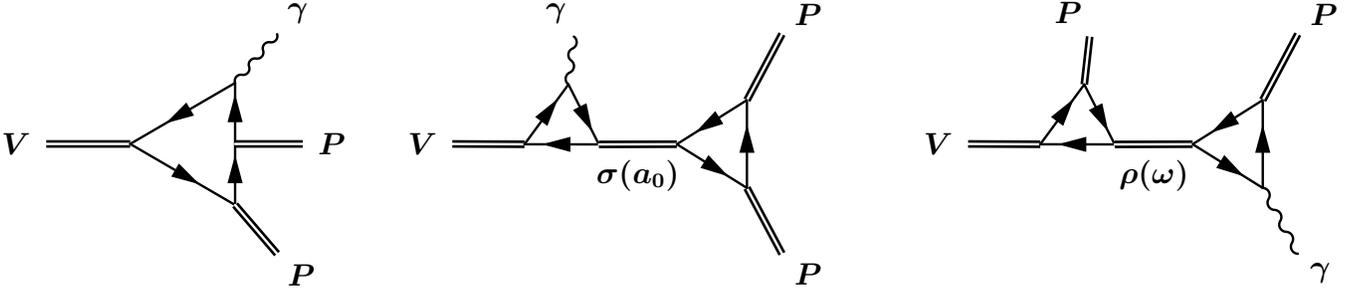}} \caption{\label{fig:VPPG} Diagrams
contributing to the amplitudes of the decays of vector mesons into pair pseudoscalar
mesons and photon.}
\end{figure*}

The processes of rare decays of vector mesons are described by three types of diagrams
(see fig. \ref{fig:VPPG}): quark box and diagrams with intermediate scalar and vector
mesons.

Let us consider vertices that are contained in the resonance diagrams and calculate the
corresponding physical processes.

The amplitude of the processes of strong decay of scalar mesons $a_0(980)\to \eta \pi$ and
$\sigma \to \pi \pi$ expressed through the divergent integral $I^{\Lambda}_2(m)$ and has the form
\begin{eqnarray}
g_{\sigma\pi\pi}&=& 8 \cos\bar\phi  g_{\sigma_u} g_{\pi}^2 I^{\Lambda}_2(m_u) = \cos\bar\phi \frac{2m_u^2Z^{1/2}}{f_\pi},\\
g_{a_0\eta\pi}  &=&-8 \sin\bar\theta g_{a_0} g_{\eta_u} g_{\pi}
I^{\Lambda}_2(m_u)=-\sin\bar\theta\frac{2m_u^2Z^{1/2}}{f_\pi}.\nonumber
\end{eqnarray}
The corresponding strong decay widths of scalar mesons takes the form
\begin{eqnarray}
\Gamma_{\sigma}(s)&=&\frac{3 g_{\sigma\pi\pi}^2}{8\pi
s}\sqrt{1-\frac{4M^2_{\pi}}{s}},\quad\Gamma_{a_0}(s)=\frac{g_{a_0\eta\pi}^2}{4\pi
s}\times\\
&&\,\times\sqrt{\left(1-\frac{(M_\eta+M_\pi)^2}{s}\right)\left(1-\frac{(M_\eta-M_\pi)^2}{s}\right)}
\nonumber
\end{eqnarray}
As a result, the decay widths $a_0(980)\to \eta \pi$ and $\sigma \to \pi \pi$ are in agreement
with experiment(see table \ref{tab:decay}).

The momentum dependent width of $\rho$-meson is
\begin{eqnarray}
\Gamma_{\rho}(s)&=&\frac{ g_{\rho}^2 s}{48\pi }\left(1-\frac{4M^2_{\pi}}{s}\right)^{3/2}
\label{rhowidth}\nonumber
\end{eqnarray}

The decays $\rho(\omega) \to \pi \gamma$, $\rho(\omega) \to \eta \gamma$ are described by
anomalous triangle diagrams. The amplitudes for the processes $\rho(\omega)\to\eta(\pi)\gamma$
have the form
\begin{eqnarray}
A(V P \gamma) \ = \ C_{VPV}\frac{g_{\rho}}{8 \pi f_\pi^2} \ \epsilon_{\mu\nu\alpha\beta} \
\epsilon_1^{\mu} \epsilon_2^{\nu} \;q_1^\alpha q_2^\beta\ ,
\end{eqnarray}
where $q_1,q_2$ and $\epsilon^{1}_\mu,\epsilon_2^{\nu}$ are the momentum and the polarization
vector of vector meson and photon, respectively. The factors $C_{VPV}$ are: $C_{\rho \pi
\gamma}=e$, $C_{\rho \eta \gamma}=-3\sin{\bar \theta}e$, $C_{\omega \pi \gamma}=3e$,
$C_{\omega \eta \gamma}=-\sin{\bar \theta}e$, $C_{\rho \pi \omega}=g_\rho$, $C_{\rho \eta
\rho}=-\sin{\bar \theta}g_\rho$, $C_{\omega \eta \omega}=-\sin{\bar \theta}g_\rho$. Note that
decays of $\rho(\omega) \to \pi \gamma$ are in good agreement with experiment, whereas
$\rho(\omega) \to \eta \gamma$ are in qualitative agreement with experiment.

At the end of this section, we should like to note that the main decay of $\omega$-meson
$\omega \to 3 \pi$ is slightly above experimental data. In \cite{Volkov:1991yr} this process
is calculated using form-factor in the $\rho \pi \pi$ vertex, which leads to better agreement
with experiment. Compilation of model predictions for widths of different decay processes is
given in table \ref{tab:decay} together with experimental results.

\begin{table}
\caption{\label{tab:decay} Meson decays obtained in the model in comparison with experimental
data \cite{PDBook}.}
\begin{tabular}{|c|c|c|c|c|c|c|c|c|c|}
 \hline
          &theory& experiment  \\
 \hline
$ \rho  \to \pi\pi        $, MeV &$ 149  $&$ 149.4\pm1.0 $    \\
$ \sigma\to \pi\pi        $, MeV &$ 582  $&$ 600-1000 $    \\
$ a_0   \to \eta\pi       $, MeV &$ 105  $&$ 50-100 $    \\
$ \omega\to 3\pi          $, MeV &$ 9.05 $&$ 7.56\pm0.07 $    \\
$ \rho^0\to\pi^0\gamma    $, KeV &$ 86   $&$ 90 \pm 13 $    \\
$ \rho^0\to\eta \gamma    $, KeV &$ 69   $&$ 44 \pm 5  $   \\
$ \omega\to\pi\gamma      $, KeV &$ 771  $&$ 755^{+30}_{-26} $\\
$ \omega\to\eta\gamma     $, KeV &$ 7.68 $&$ 4.16\pm 0.46   $ \\
 \hline
\end{tabular}
\end{table}

\section{Decays $\rho(\omega) \to \pi^0(\eta) \pi^0 \gamma$}

The amplitudes for the $\rho(\omega) \to \pi^0(\eta) \pi^0 \gamma$ decay processes are
described by three types of diagrams, see fig. \ref{fig:VPPG}. Possible combinations of
intermediate states for the diagrams with meson exchanges are the following:
\begin{itemize}
\item $\rho \to$  ($\sigma \gamma$  and  $\omega \pi^0$) $\to \pi^0 \pi^0 \gamma$

\item $\omega \to$  ($\sigma \gamma$  and  $\rho   \pi^0$) $\to \pi^0 \pi^0 \gamma$

\item $\rho \to$  ($a_0 \gamma$  and  $\omega \pi^0$ and $\rho \eta$) $\to \eta  \pi^0 \gamma$

\item $\omega \to$  ($a_0 \gamma$ and $\rho \pi^0$ and $\omega \eta$) $\to \eta  \pi^0 \gamma$
\end{itemize}

Let us consider these contributions in detail.

The quark box diagram has the form \cite{Kreopalov:1982js}
\begin{eqnarray}
A^{\mu\nu}_\mathrm{box} &=& C_{VPP\gamma}\frac{5 e g_\rho}{(6\pi f_\pi)^2}(g^{\mu\nu} (p \cdot
q_1) -p^\nu q_3^\mu), \nonumber
\end{eqnarray}
where $p,q_3$ are the momentum of vector meson and photon; factors $C_{VPP\gamma}$ are
$C_{\rho\pi^0\pi^0\gamma}=1$, $C_{\omega\pi^0\pi^0\gamma}=1/3$,
$C_{\rho\eta\pi^0\gamma}=-\sin{\bar \theta}/3$, $C_{\omega\eta\pi^0\gamma}=-\sin{\bar
\theta}$.

Scalar exchange diagrams have the same tensor structure as the quark box. Therefore it is
convenient to combine these amplitudes
\begin{eqnarray}
A^{\mu\nu}_\mathrm{box+scalar} &=& C_{VPP\gamma}\frac{5 e g_\rho}{(6\pi f_\pi)^2}(g^{\mu\nu}
(p \cdot q) -p^\nu q^\mu)\times\nonumber\\
&&\times\left(1-\frac{4 m_u^2 C_\mathcal{S}}{m_\mathcal{S}^2-s-i m_\mathcal{S}
\Gamma_\mathcal{S}(s)}\right),
\end{eqnarray}
where $\mathcal{S}$ means scalar $\sigma$ or $a_0(980)$ meson; the factor $C_\mathcal{S}$ is
$C_{a_0}=1$ and $C_\sigma=\cos{\bar \phi}$; $s=(p-q_3)^2=(q_1+q_2)^2$, where $q_1$ and $q_2$
are the momentum of pseudoscalar mesons.

The amplitudes with the vector meson ($\rho,\omega$) exchanges consists of two quark triangles
of anomalous type and the vector meson propagator. It gives the following contributions:
\begin{eqnarray}
&&A^{\mu\nu}_\mathrm{vector} = \frac{g_{\rho}}{(8 \pi f_\pi^2)^2}
\epsilon_{\mu\delta\alpha\beta} \epsilon_{\gamma\nu\tau\lambda}
g^{\delta\gamma} p^\alpha q_3^\lambda  \times \\
&&\times\biggl[ \frac{C_{VP_1V^\prime} C_{V^\prime P_2\gamma}l_1^\beta
l_1^\tau}{m_{V^\prime}^2-l_1^2-i m_{V^\prime}
\Gamma_{V^\prime}(l_1^2)}+\frac{C_{VP_1V^{\prime\prime}} C_{V^{\prime\prime} P_2\gamma}
l_2^\beta l_2^\tau}{m_{V^{\prime\prime}}^2-l_2^2-i m_{V^{\prime\prime}}
\Gamma_{V^{\prime\prime}}(l_2^2)} \biggr],\nonumber
\end{eqnarray}
where $V^\prime,V^{\prime\prime}$ are intermediate mesons and $l_1=p-q_1$, $l_2=p-q_2$. The
possible combinations of factors for different decays are
\begin{itemize}
\item $\rho \to \pi^0 \pi^0 \gamma$
\begin{eqnarray}
C_{VP_1V^\prime}=C_{VP_1V^{\prime\prime}}=C_{\rho \pi \omega},\, C_{V^\prime
P_2\gamma}=C_{V^{\prime\prime} P_2\gamma}=C_{\omega \pi \gamma}\nonumber
\end{eqnarray}
\item $\omega \to \pi^0 \pi^0 \gamma$
\begin{eqnarray}
C_{VP_1V^\prime}=C_{VP_1V^{\prime\prime}}=C_{\rho  \pi \omega},\, C_{V^\prime
P_2\gamma}=C_{V^{\prime\prime} P_2\gamma}=C_{\rho  \pi \gamma}\nonumber
\end{eqnarray}
\item $\rho \to \eta  \pi^0 \gamma$
\begin{eqnarray}
C_{VP_1V^\prime}&=&C_{\rho \pi \omega},\,C_{VP_1V^{\prime\prime}}=C_{\rho \eta \rho},\nonumber\\
C_{V^\prime P_2\gamma}&=&C_{\omega \eta \gamma},\,C_{V^{\prime\prime} P_2\gamma}=C_{\omega \pi
\gamma}\nonumber
\end{eqnarray}

\item $\omega \to \eta \pi^0 \gamma$
\begin{eqnarray}
C_{VP_1V^\prime}&=&C_{\rho  \pi \omega},\, C_{VP_1V^{\prime\prime}}=C_{\omega \eta \omega},\nonumber\\
C_{V^\prime P_2\gamma}&=&C_{\rho  \pi \gamma},\, C_{V^{\prime\prime} P_2\gamma}=C_{\omega \pi
\gamma}\nonumber
\end{eqnarray}

\end{itemize}

The width of $\rho$ meson is given by eq. (\ref{rhowidth}) and the width of $\omega$ meson is
ignored.

The decay widths have the form
\begin{eqnarray}
\Gamma_{V\pi^0P\gamma} &=&\frac{1}{192\pi^3M_V}\int\limits_{0}^{E_\gamma^\mathrm{max}}
dE_\gamma\int\limits_{E_\pi^\mathrm{min}}^{E_\pi^\mathrm{max}} dE_\pi |A_{V\pi^0P\gamma}|^2
\nonumber,
\end{eqnarray}
\begin{eqnarray}
&&E_\gamma^\mathrm{max}=\frac{M_V^2-(M_\pi+M_P)^2}{2M_V}, \\
&&E_\pi^\mathrm{min,max}=\frac{1}{2M_V(2E_\gamma-M_V)}\times\nonumber \\
&&\times\biggl[(M_V-E_\gamma)(M_V(2E_\gamma-M_V)+(M_\pi^2-M_P^2))\pm E_\gamma\times\nonumber \\
&&\times (M_V(2E_\gamma-M_V)+(M_\pi-M_P)^2)^{1/2}\times\nonumber\\
&&\times (M_V(2E_\gamma-M_V)+(M_\pi+M_P)^2)^{1/2}\biggr]\nonumber
\end{eqnarray}
where $E_\gamma$, $E_\pi$ are energies of photon and pion, $P$ means pion or $\eta$-meson(in case of pion additional
factor $1/2$ required).

The resulting branching ratios for rare processes are given in table \ref{tab:parrat}.

\begin{table}
\caption{\label{tab:parrat}Different contributions to the branching ratios for the processes
of rare decay of vector mesons. }
\begin{tabular}{|c|c|c|c|c|c|c|c|c|c|}
 \hline
process &box+scalar &      vector &       interference  &       total \\
 \hline
$\rho\rightarrow\pi^0\pi^0\gamma$, $10^{-5}$ &2.36   & 1.37  & 0.27 &   4.0 \\
$\omega\rightarrow\pi^0\pi^0\gamma$, $10^{-5}$&4.97   & 2.86  & 0.47 &   8.3 \\
$\rho\rightarrow\eta  \pi^0\gamma$, $10^{-9}$ &0.56   & 0.85  & 0.23 &   1.64 \\
$\omega\rightarrow\eta  \pi^0\gamma$, $10^{-7}$&1.22   & 1.92  & 0.51 &   3.65 \\
 \hline
\end{tabular}
\end{table}

\begin{figure*}
\resizebox{86mm}{!}{\includegraphics{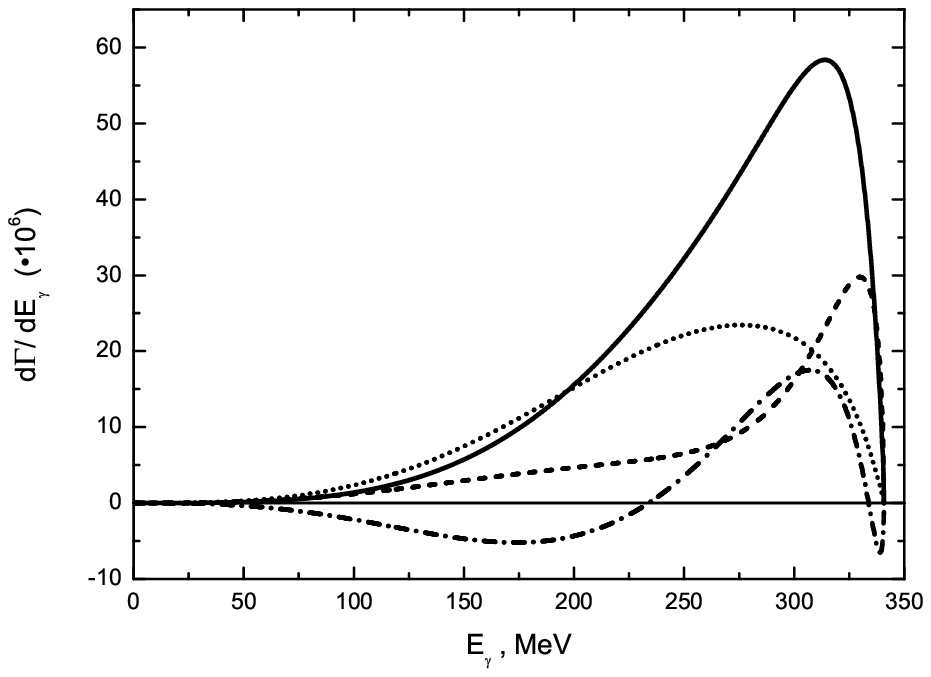}} \caption{\label{fig:rpipig} Photon spectra
for width of the decay $\rho\to \pi^0 \pi^0\gamma$. The different contribution are shown:
box+scalar meson contributions(dots), vector meson contributions(short dash),
interference(dash-dot) and total spectra(continuous line).}
\end{figure*}

\begin{figure*}
\resizebox{86mm}{!}{\includegraphics{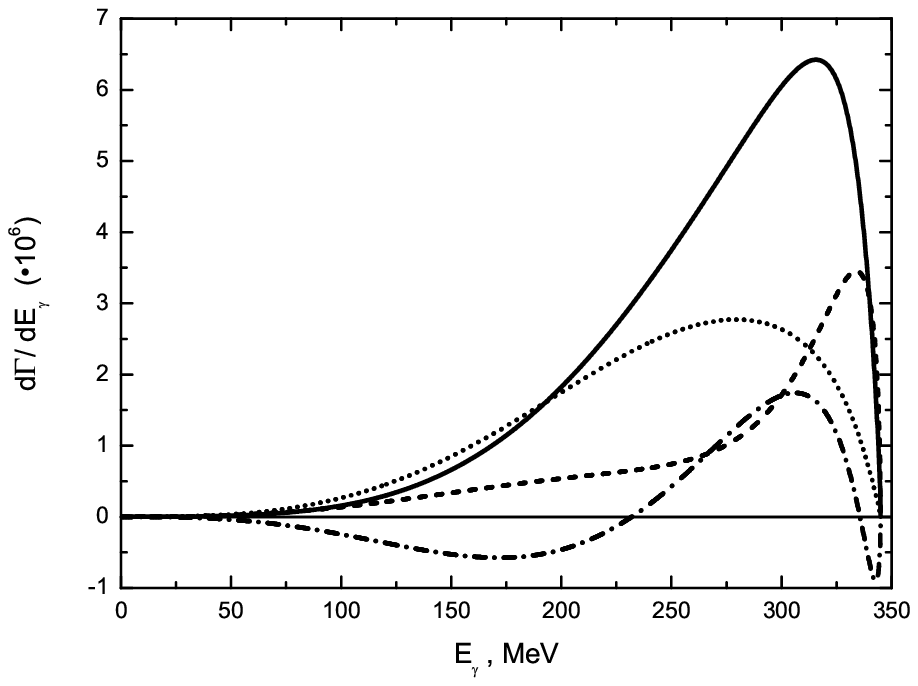}} \caption{\label{fig:opipig} Photon spectra
for width of the decay $\omega\to \pi^0 \pi^0\gamma$. The different contribution are shown:
box+scalar meson contributions(dots), vector meson contributions(short dash),
interference(dash-dot) and total spectra(continuous line).}
\end{figure*}

\begin{figure*}
\resizebox{86mm}{!}{\includegraphics{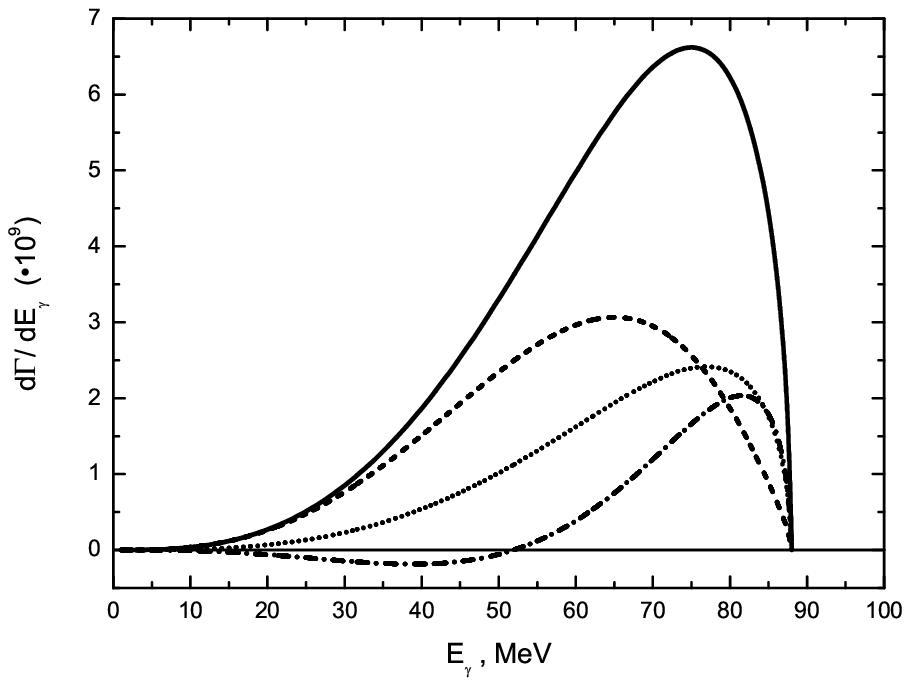}} \caption{\label{fig:retpig} Photon spectra
for width of the decay $\rho\to \eta  \pi^0\gamma$. The different contribution are shown:
box+scalar meson contributions(dots), vector meson contributions(short dash),
interference(dash-dot) and total spectra(continuous line).}
\end{figure*}

\begin{figure*}
\resizebox{86mm}{!}{\includegraphics{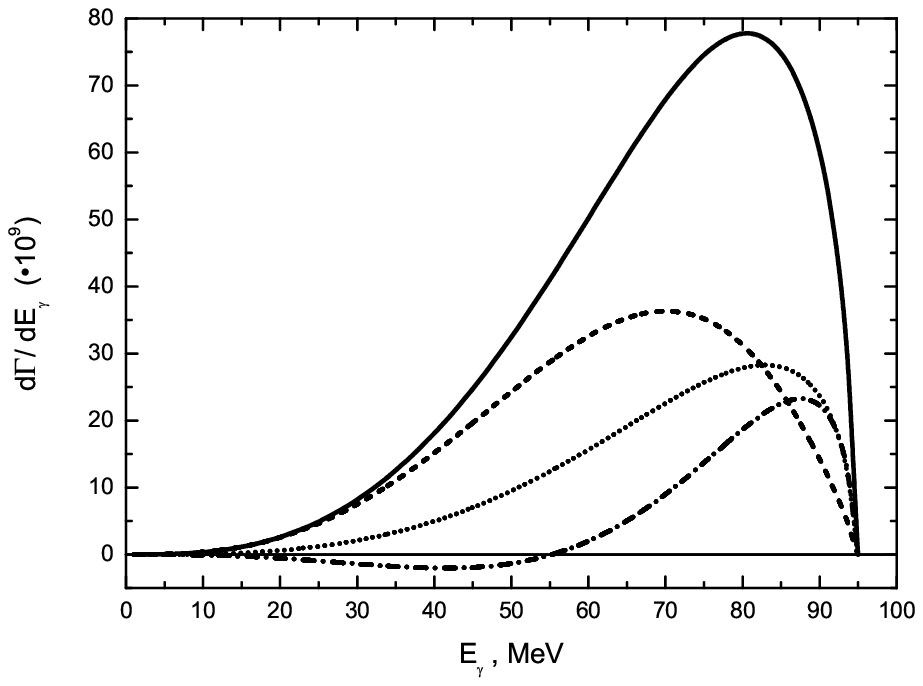}} \caption{\label{fig:oetpig}Photon spectra
for width of the decay $\omega\to \eta  \pi^0\gamma$. The different contribution are shown:
box+scalar meson contributions(dots), vector meson contributions(short dash),
interference(dash-dot) and total spectra(continuous line).}
\end{figure*}

The photon spectra for widths the of the decays $\rho\to \pi^0 \pi^0\gamma$, $\omega\to \pi^0
\pi^0\gamma$, $\rho\to \eta  \pi^0\gamma$, $\omega\to \eta  \pi^0\gamma$ are shown in figs.
\ref{fig:rpipig}, \ref{fig:opipig}, \ref{fig:retpig}, \ref{fig:oetpig}, respectively .

\section{Discussion and conclusion}

The performed calculations show that the decays widths of the processes $\rho(\omega)\to \pi^0
\pi^0\gamma$ are in satisfactory agreement with existing experimental data. The branching
ratio for the process $\omega \to \eta \pi^0 \gamma$ also does not contradict the existing
experimental limit. Comparing our result with predictions of different theoretical models we
can see that decays width with $\eta$-meson is significantly different. Therefore, obtaining
the experimental data for these processes is a very topical problem. We would like to
emphasize that in our calculation of rare processes in the framework of the standard NJL model
no additional parameters are used.

Note, that in difference with poor experimental data on decays
$\rho(\omega)\to\eta\pi^0\gamma$ the situation is much better for decays of $\phi$-meson where
rich and accurate experimental information exists. In future we plan to investigate different
decays of $\phi$-meson. We hope to obtain reasonable results because predictions of the NJL
model \cite{Volkov:1986ec,Volkov:1993jw,Ebert:1994mf} for the $\phi$-meson mass and the main
strong decay into two kaons are in satisfactory agreement with experiment \cite{PDBook}.

\begin{acknowledgments}
The authors thank A. E. Dorokhov and V. L. Yudichev for useful discussions. The authors
acknowledge the support of the Russian Foundation for Basic Research, under contract
05-02-16699.
\end{acknowledgments}



\end{document}